\documentclass{elsart}

\usepackage{graphics}
\usepackage{graphicx,float}
\usepackage{epsfig}
\usepackage{amsmath,amsfonts,amsgen}
\usepackage{dcolumn}
\usepackage[all]{xy}
\usepackage{makeidx,color}
\usepackage{color}

\newcolumntype{d}[1]{D{.}{\cdot}{#1}}
\newcolumntype{.}{D{.}{.}{-1}}

\newcommand\tabcaption{\def\@captype{table}\caption}
\newcommand{\be}{\begin{eqnarray*}}
\newcommand{\ee}{\end{eqnarray*}}
\newcommand{\ibe}{\begin{eqnarray}}
\newcommand{\iee}{\end{eqnarray}}

\def\3hf{\frac{3}{2}}

\def\np1{n+1}

\def\ip3hf{i+\frac{3}{2}}
\def\jp3hf{j+\frac{3}{2}}

\begin{document}
\begin{frontmatter}

\title{A new fractional operator of variable order: application in the description of anomalous diffusion}

\author{Xiao-Jun Yang$^{1,2,\ast}$ and J. A. Tenreiro Machado$^{3}$}
\corauth[cor1]{E-Mails: dyangxiaojun@163.com (X. J. Yang); jtm@isep.ipp.pt (J. A. T. Machado)}

\address{$^{1}$ School of Mechanics and Civil Engineering, China University of Mining and Technology, Xuzhou 221116, China}
\address{$^{2}$ State Key Laboratory for Geomechanics and Deep Underground Engineering,
China University of Mining and Technology, Xuzhou 221116, China}
\address{
$^{3}$ Institute of Engineering, Polytechnic of Porto, Department of Electrical
Engineering, Rua Dr. Antonio Bernardino de Almeida, 4249-015 Porto, Portugal}

\begin{abstract}
In this paper, a new fractional operator of variable order with
the use of the monotonic increasing function is proposed in sense of Caputo type.
The properties in term of the Laplace and Fourier
transforms are analyzed and the results for the anomalous diffusion equations of variable
order are discussed. The new
formulation is efficient in modeling a class of concentrations
in the complex transport process.
\end{abstract}


\end{frontmatter}

\section{Introduction}

Fractional-order derivatives (FOD) of the Riemann-Liouville and Caputo
types with respect to the power-law-function kernel \cite{1,2,3,4} are important
for developing mathematical models in the areas of the
control, nuclear physics, electrical circuits, signal processing,
economy and biology. FOD were used to describe anomalous diffusion (AD) problems \cite{5,6,7}. For
example, the Riemann-Liouville-type model of the AD was considered in \cite{8,9},
the Caputo-type model of the AD was reported in \cite{10}, and the AD equation with a
generalized FOD of Riemann--Liouville type was proposed in \cite{11}.

POD using the function kernel of the exponential type
were reported in \cite{12,13,14,15}. The applications of the FOD to model the surface
of shallow water (SSW) \cite{16}, heat transfer \cite{17} and linear
viscoelasticity \cite{18} were  also studied. Moreover, FOD involving the
generalized Mittag-Leffler function were proposed in \cite{19,20} and their
application in chaos was presented in \cite{21}.

More recently, a new FOD of the differentiable function with respect to the monotonic
increasing function was proposed in the sense of the Caputo type \cite{22}. The FOD
of the differentiable function involving the monotonic increasing function was
discussed in the sense of the Riemann-Liouville type \cite{22}. To our best
knowledge, FOD of the differentiable function of variable order with respect to the
monotonic increasing function (MIF) in the sense of the Caputo type has not been reported.
Motivated by this idea \cite{27,28}, in this paper the new FOD is applied in the modeling of
anomalous diffusion problems.

The paper is organized as follows. In Section 2, the
concepts of FODs of Caputo and Riemann-Liouville types with the use of the MIF
are presented. In Section 3, the results for the anomalous
diffusion models (ADMs) are discussed. Finally, in Section 4 the
conclusion are drawn.

\section{A new FOD with respect to the MIF in the sense of the Caputo
type }

In this section, the concept of the FODs of the differentiable function of
variable order in the sense of the Caputo
type is presented. We start by recalling the FODs of the differentiable
function within the MIF in the sense of the Caputo and
Riemann-Liouville types.

\subsection{The Almeida FOD of constant order with respect to the MIF}

Suppose that $T$ is the interval $\infty \le a<b\le +\infty $,
$\varphi \in C^1\left( T \right)$ with $\varphi ^{\left( 1 \right)}\left( t
\right)\ne 0$ and $\psi \in C^1\left( T \right)$ for $\forall t\in T$.

The left and right Almeida FOD of the function $\psi \left( t \right)$ of order $\xi$ $(0<\xi <1)$
in the sense of the Caputo type are given by (see \cite{22}):
\begin{equation}
\label{eq1}
{}^CD_{a^{\rm{ + }} }^{\left( {\xi ,\varphi } \right)} \psi \left( t \right) = \frac{1}{{\Gamma \left( {1 - \xi } \right)}}\int\limits_{a^{\rm{ + }} }^t {\frac{{\psi _\varphi ^{\left( 1 \right)} \left( \tau  \right)}}{{\left( {\varphi \left( t \right) - \varphi \left( \tau  \right)} \right)^\xi  }}d\tau },
\end{equation}
\begin{equation}
\label{eq2}
{ }^CD_{a^-}^{\left( {\xi ,\varphi } \right)} \psi \left( t
\right)=\frac{-1}{\Gamma \left( {1-\xi } \right)}\int\limits_{a^{\rm{ - }} }^t {\frac{\psi
_\varphi ^{\left( 1 \right)} \left( \tau \right)}{\left( {\varphi \left(
\tau \right)-\varphi \left( t \right)} \right)^\xi }d\tau } ,
\end{equation}
respectively, where $\varphi \left( t \right)$ is the MIF and $\Gamma
\left( \cdot \right)$ is the Gamma function (GF).

The Mattag-Leffler function is defined as \cite{2,4,28}:
\begin{equation}
\label{eq3}
E_\xi \left( {\eta ^\xi } \right)=\sum\limits_{i=0}^\infty {\eta ^{i\xi
}/\Gamma \left( {1+i\xi } \right)} _{.}
\end{equation}
The properties of the Almeida FOD \cite{22} are as listed in Table 1.

\begin{table}[htbp]
\begin{center}
\caption{The properties of the Almeida FOD where $\upsilon $ is real number. \label{T1}}
\begin{tabular}{ll}
\hline
Functions&
Almeida FOD
 \\
\hline
$\frac{\left( {\varphi \left( t \right)-\varphi \left( {a^+} \right)} \right)^\upsilon }{\Gamma \left( {1+\upsilon } \right)}$&
$\frac{\left( {\varphi \left( t \right)-\varphi \left( {a^+} \right)} \right)^{\upsilon -\xi }}{\Gamma \left( {1+\upsilon -\xi } \right)}$\\

$\frac{\left( {\varphi \left( {a^-} \right)-\varphi \left( t \right)} \right)^\upsilon }{\Gamma \left( {1+\upsilon } \right)}$&
$\frac{\left( {\varphi \left( {a^{-1}} \right)-\varphi \left( t \right)} \right)^{\upsilon -\xi }}{\Gamma \left( {1+\upsilon -\xi } \right)}$\\

$E_\xi \left( {\left( {\varphi \left( t \right)-\varphi \left( {a^+} \right)} \right)^\xi } \right)$&
$E_\xi \left( {\left( {\varphi \left( t \right)-\varphi \left( {a^+} \right)} \right)^\xi } \right)$
 \\
$E_\xi \left( {\left( {\varphi \left( {a^-} \right)-\varphi \left( t \right)} \right)^\xi } \right)$&
$E_\xi \left( {\left( {\varphi \left( {a^-} \right)-\varphi \left( t \right)} \right)^\xi } \right)$
 \\
\hline
\end{tabular}\\[10pt]
\end{center}
\end{table}

The left and right FOD of the function  $\psi \left( t \right)$ of order $\xi (0<\xi <1) $ within the MIF in the
sense of the Riemann-Liouville type are given by [22]:
\begin{equation}
\label{eq4}
{}^{RL}D_{a^{\rm{ + }} }^{\left( {\xi ,\varphi } \right)} \psi \left( t \right) = \frac{1}{{\Gamma \left( {1 - \xi } \right)}}\frac{1}{{\varphi ^{\left( 1 \right)} \left( t \right)}}\frac{d}{{dt}}\int\limits_{a^{\rm{ + }} }^t {\frac{{\varphi _\varphi ^{\left( 1 \right)} \left( \tau  \right)\psi \left( \tau  \right)}}{{\left( {\varphi \left( t \right) - \varphi \left( \tau  \right)} \right)^\xi  }}d\tau },
\end{equation}
\begin{equation}
\label{eq5}
{ }^{RL}D_{a^-}^{\left( {\xi ,\varphi } \right)} \psi \left( t
\right)=\frac{-1}{\Gamma \left( {1-\xi } \right)}\frac{1}{\varphi ^{\left( 1
\right)}\left( t \right)}\frac{d}{dt}\int\limits_{a^{\rm{ - }} }^t {\frac{\psi _\varphi
^{\left( 1 \right)} \left( \tau \right)}{\left( {\varphi \left( \tau
\right)-\varphi \left( t \right)} \right)^\xi }d\tau },
\end{equation}
respectively.

The relationship between Eqs. (\ref{eq1}) and (\ref{eq4}) and further details
about Eqs. (\ref{eq1}) and (\ref{eq2}) can be seen in \cite{2,22,23,24}.

\subsection{The new FOD of variable order with respect to the MIF}

Suppose that $T$ is the interval $\infty \le a<b\le +\infty $, $\varphi \in
C^1\left( T \right)$ with $\varphi ^{\left( 1 \right)}\left( t \right)\ne
0$, $\psi \in C^1\left( T \right)$ and $0<\xi \left( t \right)<1$ for
$\forall t\in T$.

The left and right $\varphi $-FOD of the function $\psi \left(
t \right)$ of order $\xi \left( t \right)$ are defined by:
\begin{equation}
\label{eq6}
{}^CD_{a^{\rm{ + }} }^{\left( {\xi \left( t \right),\varphi } \right)} \psi \left( t \right) = \frac{1}{{\Gamma \left( {1 - \xi \left( t \right)} \right)}}\int\limits_{a^{\rm{ + }} }^t {\frac{{\psi _\varphi ^{\left( 1 \right)} \left( \tau  \right)}}{{\left( {\varphi \left( t \right) - \varphi \left( \tau  \right)} \right)^{\xi \left( t \right)} }}d\tau },
\end{equation}
\begin{equation}
\label{eq7}
{ }^CD_{a^-}^{\left( {\xi \left( t \right),\varphi } \right)} \psi \left( t
\right)=\frac{-1}{\Gamma \left( {1-\xi \left( t \right)}
\right)}\int\limits_{a^{\rm{ - }} }^t {\frac{\psi _\varphi ^{\left( 1 \right)} \left( \tau
\right)}{\left( {\varphi \left( \tau \right)-\varphi \left( t \right)}
\right)^{\xi \left( t \right)}}d\tau } ,
\end{equation}
respectively, where $\varphi \left( t \right)$ is the MIF.

In a similar manner, the properties of the new $\varphi $-FOD are shown in Table 2.

\begin{table}[htbp]
\begin{center}
\caption{The properties of the new $\varphi $-FOD of order $\xi
\left( t \right)$. \label{T2}}
\begin{tabular}{ll}
\hline
Functions&
new $\varphi $-FOD
 \\
\hline
$\frac{\left( {\varphi \left( t \right)-\varphi \left( {a^+} \right)} \right)^{\xi \left( t \right)}}{\Gamma \left( {1-\xi \left( t \right)} \right)}$&
$\frac{\left( {\varphi \left( t \right)-\varphi \left( {a^+} \right)} \right)^{\upsilon -\xi \left( t \right)}}{\Gamma \left( {1+\upsilon -\xi \left( t \right)} \right)}$\\

$\frac{\left( {\varphi \left( {a^-} \right)-\varphi \left( t \right)} \right)^{\xi \left( t \right)}}{\Gamma \left( {1-\xi \left( t \right)} \right)}$&
$\frac{\left( {\varphi \left( {a^-} \right)-\varphi \left( t \right)} \right)^{\upsilon -\xi \left( t \right)}}{\Gamma \left( {1+\upsilon -\xi \left( t \right)} \right)}$
 \\

$E_\xi \left( {\left( {\varphi \left( t \right)-\varphi \left( {a^+} \right)} \right)^\xi } \right)$&
$E_\xi \left( {\left( {\varphi \left( t \right)-\varphi \left( {a^+} \right)} \right)^\xi } \right)$\\

$E_\xi \left( {\left( {\varphi \left( {a^-} \right)-\varphi \left( t \right)} \right)^\xi } \right)$&
$E_\xi \left( {\left( {\varphi \left( {a^-} \right)-\varphi \left( t \right)} \right)^\xi } \right)$
 \\
\hline
\end{tabular}\\[10pt]
\end{center}
\end{table}

Let $0<\xi \left( {\sigma ,t} \right)<1$. The left and right $\varphi $-FOD of the
function $\psi \left( {\sigma,t} \right)$ of two-variable order $\xi \left( {\sigma
,t} \right)$ are defined by:
\begin{equation}
\label{eq8}
{}^CD_{a^{\rm{ + }} }^{\left( {\xi \left( {\sigma ,t} \right),\varphi } \right)} \psi \left( {\sigma ,t} \right) = \frac{1}{{\Gamma \left( {1 - \xi \left( {\sigma ,t} \right)} \right)}}\int\limits_{a^{\rm{ + }} }^t {\frac{{\psi _\varphi ^{\left( 1 \right)} \left( {\sigma ,\tau } \right)}}{{\left( {\varphi \left( t \right) - \varphi \left( \tau  \right)} \right)^{\xi \left( {\sigma ,t} \right)} }}d\tau },
\end{equation}
\begin{equation}
\label{eq9}
{ }^CD_{a^-}^{\left( {\xi \left( {\sigma ,t} \right),\varphi } \right)} \psi
\left( {\sigma ,t} \right)=\frac{-1}{\Gamma \left( {1-\xi \left( {\sigma ,t}
\right)} \right)}\int\limits_{a^{\rm{ - }} }^t {\frac{\psi _\varphi ^{\left( 1 \right)}
\left( {\sigma ,\tau } \right)}{\left( {\varphi \left( \tau \right)-\varphi
\left( t \right)} \right)^{\xi \left( {\sigma ,t} \right)}}d\tau } ,
\end{equation}
respectively, where $\varphi \left( t \right)$ is the MIF.

Similarly, the properties of the new $\varphi $-FOD of the functions of
two-variable order $\xi \left( {\sigma ,t} \right)$ are listed in Table 3.

\begin{table}[htbp]
\begin{center}
\caption{The properties of the new $\varphi $-FOD of the functions of two-variable
order $\xi \left( {\sigma ,t} \right)$ where $\upsilon $ is real number. \label{T3}}
\begin{tabular}{ll}
\hline
Functions&
new $\varphi $-FOD
 \\
\hline
$\frac{\left( {\varphi \left( t \right)-\varphi \left( {a^+} \right)} \right)^{\xi \left( {\sigma ,t} \right)}}{\Gamma \left( {1-\xi \left( {\sigma ,t} \right)} \right)}$&
$\frac{\left( {\varphi \left( t \right)-\varphi \left( {a^+} \right)} \right)^{\upsilon -\xi \left( {\sigma ,t} \right)}}{\Gamma \left( {1+\upsilon -\xi \left( {\sigma ,t} \right)} \right)}$\\

$\frac{\left( {\varphi \left( t \right)-\varphi \left( {a^-} \right)} \right)^{\xi \left( {\sigma ,t} \right)}}{\Gamma \left( {1-\xi \left( {\sigma ,t} \right)} \right)}$&
$\frac{\left( {\varphi \left( t \right)-\varphi \left( {a^-} \right)} \right)^{\upsilon -\xi \left( {\sigma ,t} \right)}}{\Gamma \left( {1+\upsilon -\xi \left( {\sigma ,t} \right)} \right)}$
 \\

$E_\xi \left( {\left( {\varphi \left( t \right)-\varphi \left( {a^+} \right)} \right)^{\xi \left( {\sigma ,t} \right)}} \right)$&
$E_\xi \left( {\left( {\varphi \left( t \right)-\varphi \left( {a^+} \right)} \right)^{\xi \left( {\sigma ,t} \right)}} \right)$
\\

$E_\xi \left( {\left( {\varphi \left( {a^-} \right)-\varphi \left( t \right)} \right)^{\xi \left( {\sigma ,t} \right)}} \right)$&
$E_\xi \left( {\left( {\varphi \left( {a^-} \right)-\varphi \left( t \right)} \right)^{\xi \left( {\sigma ,t} \right)}} \right)$
\\
\hline
\end{tabular}\\[10pt]
\end{center}
\end{table}

For $\varphi \left( t \right)=t$, Eqs. (\ref{eq6}) and (\ref{eq7}) are written as:
\begin{equation}
\label{eq10}
{}^CD_{a^{\rm{ + }} }^{\left( {\xi \left( t \right)} \right)} \psi \left( t \right) = \frac{1}{{\Gamma \left( {1 - \xi \left( t \right)} \right)}}\int\limits_{a^{\rm{ + }} }^t {\frac{{\psi ^{\left( 1 \right)} \left( \tau  \right)}}{{\left( {t - \tau } \right)^{\xi \left( t \right)} }}d\tau },
\end{equation}
\begin{equation}
\label{eq11}
{ }^CD_{a^-}^{\left( {\xi \left( t \right)} \right)} \psi \left( t
\right)=\frac{-1}{\Gamma \left( {1-\xi \left( t \right)}
\right)}\int\limits_{a^{\rm{ - }} }^t {\frac{\psi ^{\left( 1 \right)}\left( \tau
\right)}{\left( {\tau -t} \right)^{\xi \left( t \right)}}d\tau } ,
\end{equation}
respectively.

For $\xi \left( t \right)=\xi $, Eqs. (\ref{eq6}) and (\ref{eq7}) can be expressed as \cite{2,24}:
\begin{equation}
\label{eq12}
{}^CD_{a^{\rm{ + }} }^{\left( \xi  \right)} \psi \left( t \right) = \frac{1}{{\Gamma \left( {1 - \xi } \right)}}\int\limits_{a^{\rm{ + }} }^t {\frac{{\psi ^{\left( 1 \right)} \left( \tau  \right)}}{{\left( {t - \tau } \right)^\xi  }}d\tau },
\end{equation}
\begin{equation}
\label{eq13}
{ }^CD_{a^-}^{\left( \xi \right)} \psi \left( t \right)=\frac{-1}{\Gamma
\left( {1-\xi } \right)}\int\limits_{a^{\rm{ - }} }^t {\frac{\psi ^{\left( 1
\right)}\left( \tau \right)}{\left( {\tau -t} \right)^\xi }d\tau } ,
\end{equation}
respectively.

As a matter of fact, we notice that Eqs. (\ref{eq10}) and (\ref{eq11}) are the left and right
Caputo FODs of variable order (see \cite{28} and the references therein),
respectively. Obviously, Eqs. (\ref{eq1}) and (\ref{eq6}) are expressed in the special
forms of Eq. (\ref{eq8}). Similarly, Eqs. (\ref{eq2}) and (\ref{eq7}) can be considered as the
special forms of Eqs. (\ref{eq9}) and (\ref{eq12}) and both $\psi \left( \tau
\right)$ and $\psi _\varphi ^{\left( 1 \right)} \left( {\sigma ,\tau }
\right)$ are continuous functions.

The Laplace transform (LT) of Eq. (\ref{eq12}) for $a=0$ is
given as \cite{2}:
\begin{equation}
\label{eq14}
\widehat{L}\left[ {\frac{1}{\Gamma \left( {1-\xi } \right)}\int\limits_0^t
{\frac{\psi ^{\left( 1 \right)}\left( \tau \right)}{\left( {t-\tau }
\right)^\xi }d\tau } } \right]=s^\xi \left( {\psi \left( s
\right)-s^{-1}\psi \left( 0 \right)} \right),
\end{equation}
where $\widehat{L}$ denotes the LT operator (LTO) with respect to
$t$.

Following the steps for obtaining Eq. (\ref{eq14}), the LT of $t^{-1-\xi
\left( t \right)}\mbox{/}\Gamma \left( {-\xi \left( t \right)} \right)$,
given by Coimbra \cite{28}, takes the form:
\begin{equation}
\label{eq15}
\widehat{L}\left[ {\frac{t^{-1-\xi \left( t \right)}}{\Gamma \left( {-\xi
\left( t \right)} \right)}} \right]=s^{\xi \left( t \right)}.
\end{equation}
With the aid of Eqs. (\ref{eq14}) and (\ref{eq15}), the LT of
$t^{-1-\xi \left( {\sigma ,t} \right)}/\Gamma \left( {-\xi \left( {\sigma
,t} \right)} \right)$ is given by:
\begin{equation}
\label{eq16}
\widehat{L}\left[ {\frac{t^{-1-\xi \left( {\sigma ,t} \right)}}{\Gamma
\left( {-\xi \left( {\sigma ,t} \right)} \right)}} \right]=s^{\xi \left(
{\sigma ,t} \right)}.
\end{equation}
For $\sigma =0$, Eq. (\ref{eq16}) can be written as Eq. (\ref{eq15}).

Similarly, the LT of $\left( {\varphi \left( t \right)}
\right)^{-1-\xi \left( {\sigma ,t} \right)}/\Gamma \left( {-\xi \left(
{\sigma ,t} \right)} \right)$ is as follows:
\begin{equation}
\label{eq17}
\widehat{L}\left[ {\frac{\left( {\varphi \left( t \right)} \right)^{-1-\xi
\left( {\sigma ,t} \right)}}{\Gamma \left( {-\xi \left( {\sigma ,t} \right)}
\right)}} \right]=\left[ {\varphi _\varphi ^{\left( 1 \right)} \left( t
\right)} \right]\left( s \right)^{\xi \left( {\sigma ,t} \right)},
\end{equation}
where $\varphi ^{\left( 1 \right)}\left( t \right)$ is the operator with
respect to $t$.

For $\varphi \left( t \right)=t$, we have
\begin{equation}
\label{eq18}
\varphi _\varphi ^{\left( 1 \right)} \left( t \right)=1
\end{equation}
so that Eq. (\ref{eq17}) becomes Eq. (\ref{eq16}), that is,
\begin{equation}
\label{eq19}
\left[ {\varphi _\varphi ^{\left( 1 \right)} \left( t \right)} \right]\left(
s \right)^{\xi \left( {\sigma ,t} \right)}=\left[ 1 \right]\left( s
\right)^{\xi \left( {\sigma ,t} \right)}=s^{\xi \left( {\sigma ,t} \right)}.
\end{equation}
In this case, we have
\begin{equation}
\label{eq20}
\widehat{L}\left[ {\frac{\left( {\varphi \left( t \right)} \right)^{-1-\xi
\left( {\sigma ,t} \right)}}{\Gamma \left( {-\xi \left( {\sigma ,t} \right)}
\right)}} \right]=\left[ {\varphi _\varphi ^{\left( 1 \right)} \left( t
\right)} \right]\left( s \right)^{\xi \left( {\sigma ,t} \right)},
\end{equation}
\begin{equation}
\label{eq21}
\widehat{L}\left[ {\psi _\varphi ^{\left( 1 \right)} \left( {\sigma ,t}
\right)} \right]=\psi \left( {\sigma ,s} \right)-\left[ {\varphi _\varphi
^{\left( 1 \right)} \left( t \right)} \right]\left( s \right)^{-1}\psi
\left( {\sigma ,0} \right)
\end{equation}
such that
\begin{equation}
\label{eq22}
\begin{array}{l}
\widehat{L}\left[{{}^CD_{0^{\rm{ + }} }^{\left( {\xi \left( {\sigma ,t} \right),\varphi } \right)} \psi \left( {\sigma ,t} \right)}\right]= \left[ {\varphi _\varphi ^{\left( 1 \right)} \left( t \right)} \right]\left( s \right)^{\xi \left( {\sigma ,t} \right)} \psi \left( s \right) - \left[ {\varphi _\varphi ^{\left( 1 \right)} \left( t \right)} \right]\left( s \right)^{\xi \left( {\sigma ,t} \right) - 1} \psi \left( 0 \right),
 \end{array}
\end{equation}
where
\begin{equation}
\label{eq23}
\widehat{L}\left[ {\frac{\left( {\varphi \left( t \right)} \right)^{-\xi
\left( {\sigma ,t} \right)}}{\Gamma \left( {1-\xi \left( {\sigma ,t}
\right)} \right)}} \right]=\left[ {\varphi _\varphi ^{\left( 1 \right)}
\left( t \right)} \right]\left( s \right)^{\xi \left( {\sigma ,t} \right)}.
\end{equation}

Taking $s=j\varpi $, we have the following Fourier transforms (FT):
\begin{equation}
\label{eq26}
\widehat{F}\left[ {\frac{\left( {\varphi \left( t \right)} \right)^{-1-\xi
\left( {\sigma ,t} \right)}}{\Gamma \left( {-\xi \left( {\sigma ,t} \right)}
\right)}} \right]=\left[ {\varphi _\varphi ^{\left( 1 \right)} \left( t
\right)} \right]\left( {j\varpi } \right)^{\xi \left( {\sigma ,t} \right)},
\end{equation}
\begin{equation}
\label{eq27}
\widehat{F}\left[ {\frac{\left( {\varphi \left( t \right)} \right)^{-1-\xi
\left( {\sigma ,t} \right)}}{\Gamma \left( {-\xi \left( {\sigma ,t} \right)}
\right)}} \right]=\left[ {\varphi _\varphi ^{\left( 1 \right)} \left( t
\right)} \right]\left( {j\varpi } \right)^{\xi \left( {\sigma ,t} \right)},
\end{equation}
\begin{equation}
\label{eq28}
\widehat{F}\left[ {{}^CD_{0^{\rm{ + }} }^{\left( {\xi \left( {\sigma ,t} \right),\varphi } \right)} \psi \left( {\sigma ,t} \right)} \right] = \left[ {\varphi _\varphi ^{\left( 1 \right)} \left( t \right)} \right]\left( {j\varpi } \right)^{\xi \left( {\sigma ,t} \right)} \psi \left( {j\varpi } \right),
\end{equation}
where $\widehat{F}$ denotes the FT operator (FTO) with respect to
$t$.

Suppose that $T$ is the interval $\infty \le a<b\le +\infty $, $\varphi \in
C^{n+1}\left( T \right)$ with $\varphi ^{\left( 1 \right)}\left( t
\right)\ne 0$, $\psi \in C^{n+1}\left( T \right)$ and $n-1<\xi \left( t
\right)<n$ for $\forall t\in T$.

The left and right $\varphi $-FOD of the
function $\psi \left( t \right)$ of order $\xi \left( t \right)$ are defined
by:
\begin{equation}
\label{eq29}
{}^CD_{a^{\rm{ + }} }^{\left( {\xi \left( t \right),\varphi } \right)} \psi \left( t \right) = \frac{1}{{\Gamma \left( {n - \xi \left( t \right)} \right)}}\int\limits_{a^{\rm{ + }} }^t {\frac{{\psi _\varphi ^{\left( n \right)} \left( \tau  \right)}}{{\left( {\varphi \left( t \right) - \varphi \left( \tau  \right)} \right)^{n - \xi \left( t \right) - 1} }}d\tau },
\end{equation}
\begin{equation}
\label{eq30}
{ }^CD_{a^-}^{\left( {\xi \left( t \right),\varphi } \right)} \psi \left( t
\right)=\frac{-1}{\Gamma \left( {n-\xi \left( t \right)}
\right)}\int\limits_{a^{\rm{ - }} }^t {\frac{\psi _\varphi ^{\left( n \right)} \left( \tau
\right)}{\left( {\varphi \left( \tau \right)-\varphi \left( t \right)}
\right)^{n-\xi \left( t \right)-1}}d\tau } ,
\end{equation}
respectively, where $\varphi \left( t \right)$ is the MIF.

Let $n-1<\xi \left( {\sigma ,t} \right)<n$. The left and right $\varphi $-FOD of the
function $\psi \left( {\sigma ,t} \right)$ of two-variable order $\xi \left( {\sigma
,t} \right)$ are defined by:
\begin{equation}
\label{eq31}
{}^CD_{a^{\rm{ + }} }^{\left( {\xi \left( {\sigma ,t} \right),\varphi } \right)} \psi \left( {\sigma ,t} \right) = \frac{1}{{\Gamma \left( {n - \xi \left( {\sigma ,t} \right)} \right)}}\int\limits_{a^{\rm{ + }} }^t {\frac{{\psi _\varphi ^{\left( n \right)} \left( {\sigma ,\tau } \right)}}{{\left( {\varphi \left( t \right) - \varphi \left( \tau  \right)} \right)^{n - \xi \left( {\sigma ,t} \right) - 1} }}d\tau },
\end{equation}
\begin{equation}
\label{eq32}
{ }^CD_{a^-}^{\left( {\xi \left( {\sigma ,t} \right),\varphi } \right)} \psi
\left( {\sigma ,t} \right)=\frac{-1}{\Gamma \left( {n-\xi \left( {\sigma ,t}
\right)} \right)}\int\limits_{a^{\rm{ - }} }^t {\frac{\psi _\varphi ^{\left( n \right)}
\left( {\sigma ,\tau } \right)}{\left( {\varphi \left( \tau \right)-\varphi
\left( t \right)} \right)^{n-\xi \left( {\sigma ,t} \right)-1}}d\tau } ,
\end{equation}
respectively, where $\varphi \left( t \right)$ is the MIF.

Similarly, the LT of Eq. (\ref{eq31}) is given by:
\begin{equation}
\label{eq33}
\widehat{L}\left[ {{}^CD_{0^{\rm{ + }} }^{\left( {\xi \left( {\sigma ,t} \right),\varphi } \right)} \psi \left( {\sigma ,t} \right)} \right]\\
= \left[ {\varphi _\varphi ^{\left( n \right)} \left( t \right)} \right]\left( s \right)^{\xi \left( {\sigma ,t} \right)} \left( {\psi \left( {\sigma ,s} \right) - \sum\limits_{\eta {\rm{ = 1}}}^{n - 1} {\left[ {\varphi _\varphi ^{\left( n \right)} \left( t \right)} \right]\left( s \right)^{ - \eta } \psi _\varphi ^{\left( {\eta  - 1} \right)} \left( {\sigma ,0} \right)} } \right),
\end{equation}
and the FT of Eq. (\ref{eq31}) by:
\begin{equation}
\label{eq34}
\widehat{F}\left[ {{}^CD_{0^{\rm{ + }} }^{\left( {\xi \left( {\sigma ,t} \right),\varphi } \right)} \psi \left( {\sigma ,t} \right)} \right] = \left[ {\varphi _\varphi ^{\left( n \right)} \left( t \right)} \right]\left( {j\varpi } \right)^{\xi \left( {\sigma ,t} \right)},
\end{equation}
where $\psi _\varphi ^{\left( 0 \right)} \left( {\sigma ,0} \right)=\psi
\left( {\sigma ,0} \right)$.

Similarly, we present the following ODE of variable order with respect to the MIF:
\begin{equation}
\label{eq35}
{}^CD_{0^{\rm{ + }} }^{\left( {\xi \left( {\sigma ,t} \right),\varphi } \right)} \psi \left( {\sigma ,t} \right) = \Lambda \psi \left( {\sigma ,t} \right)
\end{equation}
with the solution involving the MIF
\begin{equation}
\label{eq36}
\psi \left( {\sigma ,t} \right)=\Lambda E_{\xi \left( {\sigma ,t} \right)}
\left( {\left( {\varphi \left( t \right)} \right)^{\xi \left( {\sigma ,t}
\right)}} \right),
\end{equation}
where $\Lambda $ is a constant and
\[E_{\xi \left( {\sigma ,t} \right)}
\left( {\left( {\varphi \left( t \right)} \right)^{\xi \left( {\sigma ,t}
\right)}} \right)=\sum\limits_{i=0}^\infty {\left( {\varphi \left( t
\right)} \right)^{i\xi \left( {\sigma ,t} \right)}/\Gamma \left( {1+i\xi
\left( {\sigma ,t} \right)} \right)}.\]

The chart of Eq. (\ref{eq36}) for the different parameters is displayed in Figure 1.
\begin{figure}
\centering
\includegraphics[width=0.8 \linewidth]{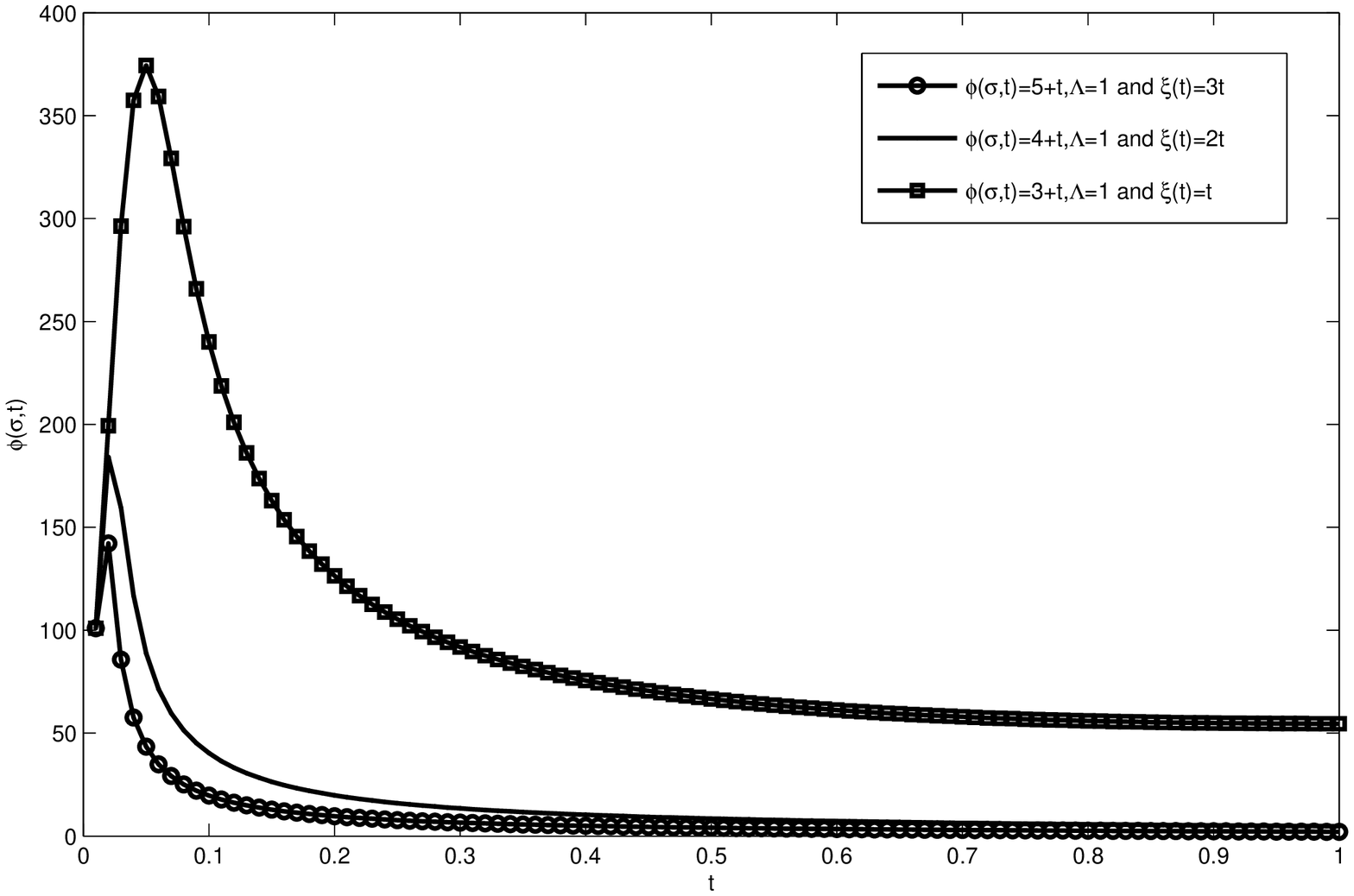}
\caption{Plot of Eq. (\ref{eq36}) for the parameters
$\varphi \left( t \right)=3+t$, $\xi \left( {\sigma ,t} \right)=t$, $\varphi \left( t \right)=4+t$,
$\xi \left( {\sigma ,t} \right)=2t$, $\varphi \left( t \right)=5+t$, $\xi \left( {\sigma ,t} \right)=3t$ and $\Lambda=1 $.}
\end{figure}

Thus, from Eq. (\ref{eq36}), we suggest the following ODE of variable order with respect to the MIF:
\begin{equation}
\label{eq37}
{}^CD_{0^{\rm{ + }} }^{\left( {\xi \left( t \right),\varphi } \right)} \psi \left( t \right) = \Lambda \psi \left( t \right)
\end{equation}
with the solution involving the MIF
\begin{equation}
\label{eq38}
\psi \left( t \right)=\Lambda E_{\xi \left( t \right)} \left( {\left(
{\varphi \left( t \right)} \right)^{\xi \left( t \right)}} \right),
\end{equation}
where
$\Lambda $ is a constant and
\[E_{\xi \left( t \right)} \left( {\left(
{\varphi \left( t \right)} \right)^{\xi \left( t \right)}}
\right)=\sum\limits_{i=0}^\infty {\left( {\varphi \left( t \right)}
\right)^{i\xi \left( t \right)}/\Gamma \left( {1+i\xi \left( t \right)}
\right)}.\]
The chart of Eq. (\ref{eq38}) for the different parameters is displayed in Figure 2.
\begin{figure}
\centering
\includegraphics[width=0.8 \linewidth]{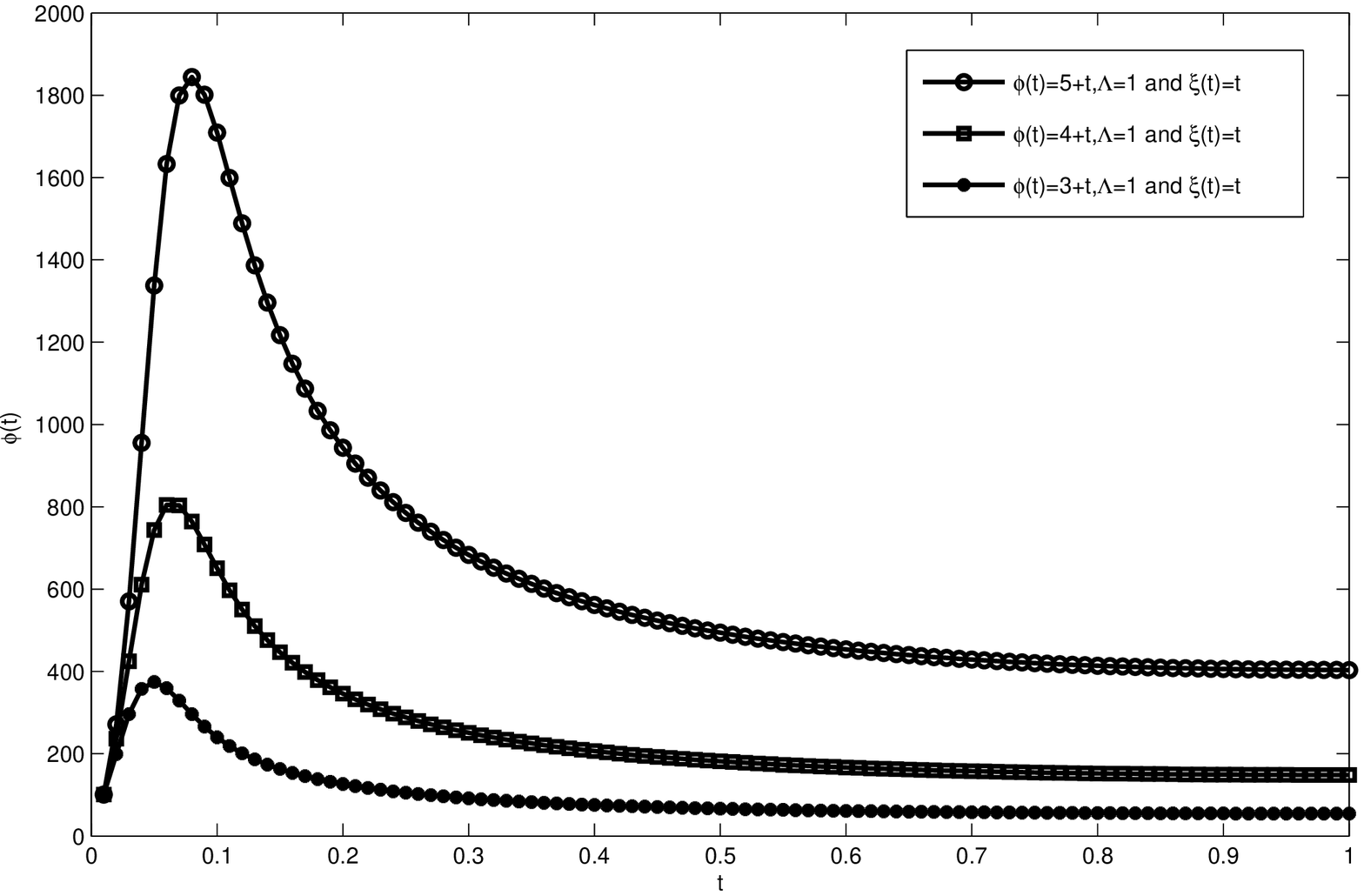}
\caption{Plot of Eq. (\ref{eq38}) for the parameters
$\varphi \left( t \right)=3+t$, $\varphi \left( t \right)=4+t$, $\varphi \left( t \right)=5+t$,
$\Lambda=1 $ and $\xi \left( {t} \right)=t$.}
\end{figure}

For $\xi \left( t \right)=\xi $, Eq. (\ref{eq38}) is in line with the result in
\cite{22}.

\section{The ADM described by $\varphi $ -FODs }

The ADMs represent the concentrations of the density of particles in the physical systems. They
play an important role in the description of the complex medium, such as
heat, electric circuit, geophysics and biological systems.

For predicting the complex behavior in the inhomogeneous media, we can
follow the ADM given by:
\begin{equation}
\label{eq39}
{}^CD_{0^{\rm{ + }} }^{\left( {\xi \left( {\sigma ,t} \right),\varphi } \right)} \psi _\rho  \left( {\sigma ,t} \right) = \varsigma \frac{{\partial ^2 \psi _\rho  \left( {\sigma ,t} \right)}}{{\partial \sigma ^2 }},
0 < \xi \left( {\sigma ,t} \right) < 1,
\end{equation}
subjected to the initial-boundary value conditions
\begin{equation}
\label{eq40}
\psi _\rho \left( {\sigma ,0} \right)=\delta \left( \sigma \right),
\end{equation}
where $\psi _\rho \left( {\sigma ,t} \right)$ is the concentrations,
$\varsigma$ $\left( {\varsigma >0} \right)_{ }$ is the diffusion constant,
and $\delta \left( \sigma \right)$ is the Dirac function.

In the above equation, the $\varphi $-FOD involving the function $\varphi \left( t
\right)$ of two-variable order $\xi \left( {\sigma ,t} \right)_{ }$ is adopted
to model the different concentrations including the constant-order (CO),
time-variable (TV), space-variable (SV) and concentration-dependent variable-order (CDVO).

Following the above consideration, we have the following:

\subsection{The COADM with respect to the monotonic increasing time-variable function}

The COADM used to describe the system of the unknown variable transport
process were
\begin{equation}
\label{eq41}
{}^CD_{0^{\rm{ + }} }^{\left( {\xi ,\varphi } \right)} \psi _\rho  \left( {\sigma ,t} \right) = \varsigma \frac{{\partial ^2 \psi _\rho  \left( {\sigma ,t} \right)}}{{\partial \sigma ^2 }}
,
0 < \xi  < 1,
\end{equation}
and
\begin{equation}
\label{eq42}
\int\limits_0^1 {o\left( \xi  \right){}^CD_{0^{\rm{ + }} }^{\left( {\xi ,\varphi } \right)} \psi _\rho  \left( {\sigma ,t} \right)d\xi }  = \varsigma \frac{{\partial ^2 \psi _\rho  \left( {\sigma ,t} \right)}}{{\partial \sigma ^2 }}
,\mbox{ }\int\limits_0^1 {o\left( \xi \right)d\xi =0} ,
\end{equation}
where $o\left( \xi \right)$ is the function considered to decelerate or
accelerate the diffusion process.

We observe that the proposed models in \cite{26} are the special cases of Eqs. (\ref{eq41}) and (\ref{eq42}). Furthermore,
from mathematical point of view, the concentration of the particle in the transport process is
also described by the COADM.
 Eq. (\ref{eq42}) is considered to present the COADM with the use of the
decelerating or accelerating processes.

\subsection{The TVADM with respect to the monotonic increasing time-variable function}

With the support of the complex behaviors of the anomalous transport
systems, we can write the following equation:
\begin{equation}
\label{eq43}
{}^CD_{a^{\rm{ + }} }^{\left( {\xi \left( t \right),\varphi } \right)} \psi _\rho  \left( {\sigma ,t} \right) = \varsigma \frac{{\partial ^2 \psi _\rho  \left( {\sigma ,t} \right)}}{{\partial \sigma ^2 }}
, 0<\xi \left( t \right)<1, \mbox{ }0\le \sigma \le \varpi ,
\end{equation}
where the initial-boundary conditions are as follows \cite{26}:
\begin{equation}
\label{eq44}
\psi _\rho \left( {\sigma ,0} \right)=\sin \left( {\frac{\sigma \pi }{\varpi
}} \right), 0\le \sigma \le \varpi,
\end{equation}
\begin{equation}
\label{eq45}
\psi _\rho \left( {0,t} \right)=\psi _\rho \left( {\varpi ,t}
\right)=0, \mbox{ }t=0.
\end{equation}
In fact, Eq. (\ref{eq43}) is the TVADM of the Caputo-type applied to
the transport process in the inhomogeneous media. When $\varphi =t$, Eq. (\ref{eq43})
is the time dependent variable-order ADM proposed in \cite{26}. When $\varphi \ne
t$, the Sun-Chen-Chen model \cite{26} is invalid for handling the generalized
complex media. However, Eq. (\ref{eq43}) can be successfully applied to model the
above problems in the different transitional regimes.

\subsection{The SVAMD with respect to the monotonic increasing time-variable function}

The SVADM in term of the monotonic increasing time-variable function is
expressed by:
\begin{equation}
\label{eq46}
{}^CD_{a^{\rm{ + }} }^{\left( {\xi \left( \sigma  \right),\varphi } \right)} \psi _\rho  \left( {\sigma ,t} \right) = \varsigma \frac{{\partial ^2 \psi _\rho  \left( {\sigma ,t} \right)}}{{\partial \sigma ^2 }}
, 0<\xi \left( \sigma \right)<1,\mbox{ }0\le \sigma \le \varpi ,
\end{equation}
where $\varpi$ is a constant, and the fractional-order-space differential operator is defined by:
\begin{equation}
\label{eq47}
{ }^CD_{a^-}^{\left( {\xi \left( \sigma \right),\varphi } \right)} \psi
\left( t \right)=\frac{1}{\Gamma \left( {n-\xi \left( \sigma \right)}
\right)}\int\limits_a^t {\frac{\psi _\varphi ^{\left( 1 \right)} \left( \tau
\right)}{\left( {\varphi \left( t \right)-\varphi \left( \tau \right)}
\right)^{\xi \left( \sigma \right)}}d\tau } .
\end{equation}
In the above result, Eq. (\ref{eq46}) is the SVADM of the Caputo-type with respect
to the monotonic increasing time-variable function and is
used to predict the transport process in the special complex media.
When $\varphi =t$, the space-dependent variable-order ADM presented in \cite{26}
is a special case of Eq. (\ref{eq43}).

\subsection{The CDVOADM with respect to the monotonic increasing time-variable function}

The CDVOADM with respect to the monotonic increasing time-variable
function is written as:
\begin{equation}
\label{eq48}
{}^CD_{a^{\rm{ + }} }^{\left( {\psi _\rho  \left( {\sigma ,t} \right),\varphi } \right)} \psi _\rho  \left( {\sigma ,t} \right) = \varsigma \frac{{\partial ^2 \psi _\rho  \left( {\sigma ,t} \right)}}{{\partial \sigma ^2 }},
0 < \xi \left( \sigma  \right) < 1, {\rm{ }}0 \le \sigma  \le \varpi,
\end{equation}
where $\varpi$ is a constant, and the fractional-order concentration-dependent differential operator is
defined by:
\begin{equation}
\label{eq49}
{}^CD_{a^{\rm{ + }} }^{\left( {\psi _\rho  \left( {\sigma ,t} \right),\varphi } \right)} \psi \left( {\sigma ,t} \right) = \frac{1}{{\Gamma \left( {1 - \xi \left( {\sigma ,t} \right)} \right)}}\int\limits_a^t {\frac{{\psi _\varphi ^{\left( 1 \right)} \left( {\sigma ,\tau } \right)}}{{\left( {\varphi \left( t \right) - \varphi \left( \tau  \right)} \right)^{\psi _\rho  \left( {\sigma ,t} \right)} }}d\tau }.
\end{equation}
We easily observe Eq. (\ref{eq49}) is the extended version of the Sun-Chen-Chen model
reported in \cite{26}, when $\varphi \left( t \right)=t$.

\section{Conclusion}

In this work, we developed a new variable-order FOD with respect to
the MIF in the sense of the Caputo type. Meanwhile, we also proposed a class of the special
functions of Mattag-Leffler type and the FT and the LT of the proposed FOD.
Finally, an ADM of $\varphi $-FOD type with respect to the monotonic
increasing time-variable function was obtained and its analogies are also discussed. The
anomalous diffusion models, as the generalized result reported in \cite{26}, are
useful for describing the transport process in the special complex media.

\end{document}